\documentclass[showpacs,showkeys,preprintnumbers,twocolumn]{revtex4}
\usepackage{amsmath}
\usepackage{hyperref}
\usepackage{graphicx}

\input{tcilatex}
\usepackage{graphicx}

\begin{document}

\title{Multi-impurity effects on the entanglement of anisotropic Heisenberg
ring XXZ under a homogeneous magnetic field}
\author{Gao Wen-Bin,Yang Guo-Hui, and Zhou Ling\ \ }
\affiliation{Department of Physics, Dalian University of Technology, Dalian 116024, PR
China}

\begin{abstract}
The effects of multi-impurity on the entanglement of anisotropic Heisenberg
ring XXZ under a homogeneous magnetic field have been studied. The
impurities make the equal pairwise entanglement in a ring compete with each
other so that the pairwise entanglement exhibits oscillation. If the
impurities are of larger couplings, both the critical temperature and
pairwise entanglement can be improved.
\end{abstract}

\maketitle

\section{\ \ Introduction}

Entanglement is not only the fabulous feature of quantum mechanics but also
very important to the quantum information processing (QIP).$^{[1]}$ In the
studies of quantum entanglement, solid state system with Heisenberg model
interaction is the simple and applicable candidates for the realization of
quantum information. Therefore, there are many works focusing mainly on the
different kinds of Heisenberg models$^{[2-21]}$ such as spin ring etc..

The impurities often exist in solid system and plays a very obvious and
important part in condensed matter physics. As a candidate of QIP, solid
system with impurity is also one of our important study object. In the
previous researches, the impurity effects on the quantum entanglement have
been studied in a three-spin system $^{[22,23]}$ and a large spin systems
under zero temperature.$^{[24]}$ However, in these works, they have just
studied single impurity.

In this paper, we will focus on studying the effects of multi- impurity on
the pairwise thermal entanglement in a ring chain. We find the impurities
make the equal pairwise entanglement in a ring compete with each other. If
impurities are of large couplings, the critical temperature and the pairwise
entanglement which coupled to the impurities can be improved. Our studying
results not only provide a standard to judge impurities but also provide a
way to enhance entanglement and critical temperature.

\section{Non-nearest Neighboring Impurity Effect}

Firstly, we investigate the multi-impurity effect when the impurities are
non-nearest neighbors shown as the Fig. 1. In the two figures, square
represents impurity qubit and round stands for normal qubit.

\FRAME{ftbpFU}{2.9308in}{2.0237in}{0pt}{\Qcb{Two configurations of spin ring
when the impurities are non-nearest neighbors. (a): qubit ring formed with
10 qubits. The 4th and 6th are two identical impurities. (b): qubit ring
formed with 10 qubits. The 4th, 6th and 8th are three identical impurities.}%
}{}{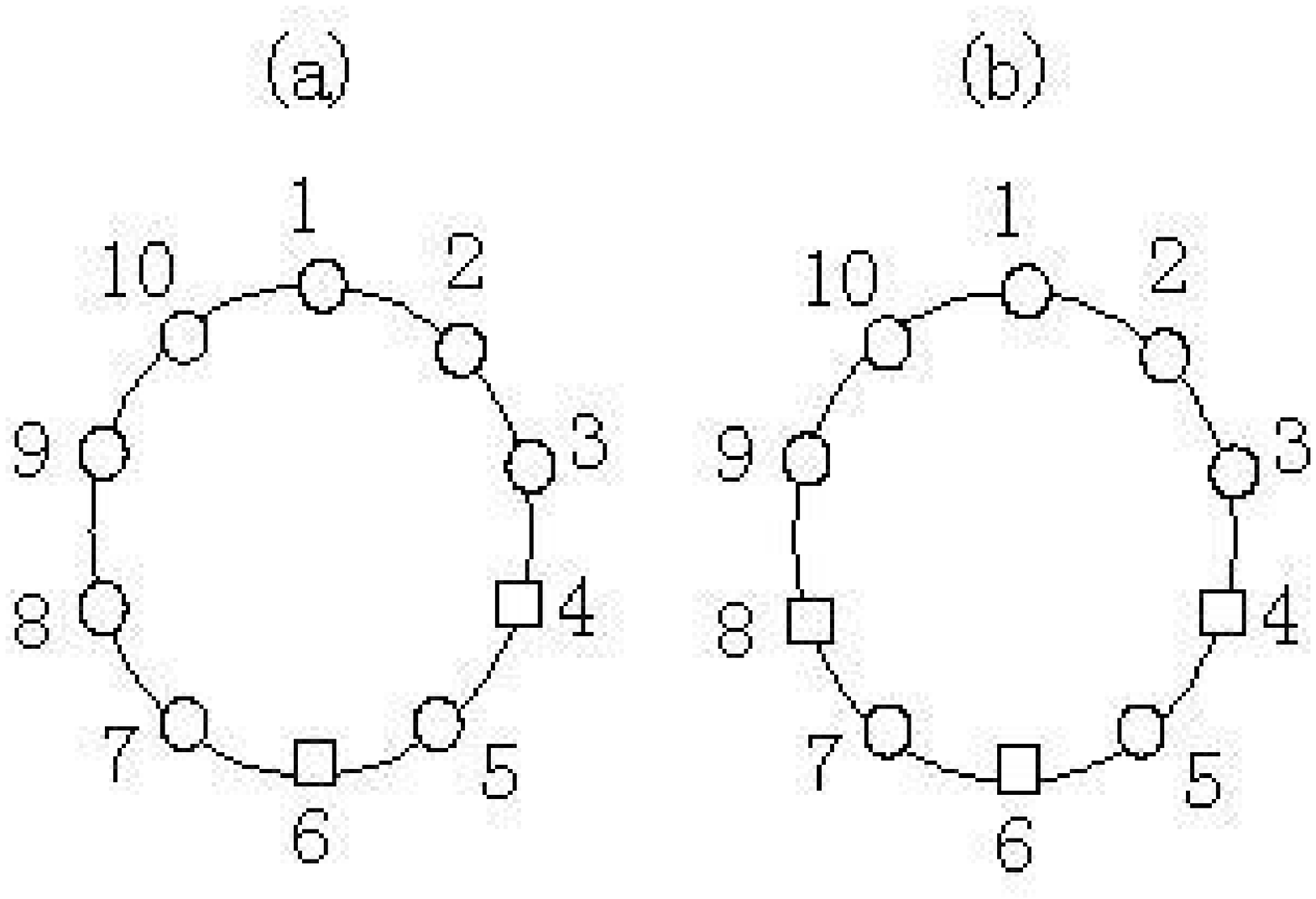}{\raisebox{-2.0237in}{\includegraphics[height=2.0237in]{fig1.eps}}}For the case of Fig. 1a, the
Hamiltonian can be written as

{\normalsize 
\begin{eqnarray}
H &=&\frac{1}{2}\sum_{i=1}^{2}[J(\sigma _{i}^{x}\sigma _{i+1}^{x}+\sigma
_{i}^{y}\sigma _{i+1}^{y})+J_{z}\sigma _{i}^{z}\sigma _{i+1}^{z}]  \notag \\
&&+\frac{1}{2}\sum_{i=7}^{N}[J(\sigma _{i}^{x}\sigma _{i+1}^{x}+\sigma
_{i}^{y}\sigma _{i+1}^{y})+J_{z}\sigma _{i}^{z}\sigma _{i+1}^{z}]  \notag \\
&&+\frac{1}{2}\sum_{i=3}^{6}[J^{^{\prime }}(\sigma _{i}^{x}\sigma
_{i+1}^{x}+\sigma _{i}^{y}\sigma _{i+1}^{y})+J_{z}^{^{\prime }}\sigma
_{i}^{z}\sigma _{i+1}^{z}]  \notag \\
&&+\frac{1}{2}\sum_{i=1}^{N}B(\sigma _{i}^{z}+\sigma _{i+1}^{z}),
\label{eq1}
\end{eqnarray}%
}

where ($\sigma _{i}^{x}$, $\sigma _{i}^{y}$, $\sigma _{i}^{z}$) are the
vector of Pauli matrices; $J$ and $J_{z}$ are the real coupling coefficients
of arbitrary nearest neighboring two qubits. We restrict the $B\geq 0$ along 
$z$ direction and $N+1=1$. We choose the parameters $B$, $J$, $J_{z}$ and $T$
are dimensionless and assume the coupling coefficients between normal qubit
and impurity one has the relation

\begin{equation}
J^{^{\prime }}=\alpha \ast J,J_{z}^{^{\prime }}=\alpha \ast J_{z},
\end{equation}%
where $\alpha $ characterizes the relative strength of the extra coupling
between the impurity and its nearest neighboring qubits.$^{[24]}$ For the
case of Fig. 1b, one can write it easily following Eq.(1). We do not give it
here any more.

As we know, for a system in equilibrium at temperature $T$, the density
operator is $\rho =(1/Z)\exp (-H/k_{B}T)$, where $Z=Tr[\exp (-H/k_{B}T)]$ is
the partition function and $k_{B}$ is Boltzman's constant. For simplicity,
we write $k_{B}=1$. The value of entanglement between two qubits can be
measured by Concurrence $C$ which is written as 
\begin{equation}
C=\max (0,2\max {\lambda _{i}}-\sum_{i=1}^{4}\lambda _{i})  \label{eq3}
\end{equation}%
$^{[25,26,27,28]}$ in which $\lambda _{i}$ is the square roots of the
eigenvalues of the matrix 
\begin{equation}
R=\rho (\sigma _{1}^{y}\otimes \sigma _{2}^{y})\rho ^{\ast }(\sigma
_{1}^{y}\otimes \sigma _{2}^{y}),  \label{eq4}
\end{equation}%
where $\rho $ is the density matrix and the symbol $\ast $ stands for
complex conjugate. The Concurrence can be calculated no matter whether $\rho 
$ is pure or mixed. In the following, we just take the pairwise entanglement
into account. We will trace over the qubits and study the reduced density
matrix of the two qubits which we are interested in. 
\begin{figure}[tbph]
{\normalsize \includegraphics[width=3.5in,height=2.5in]{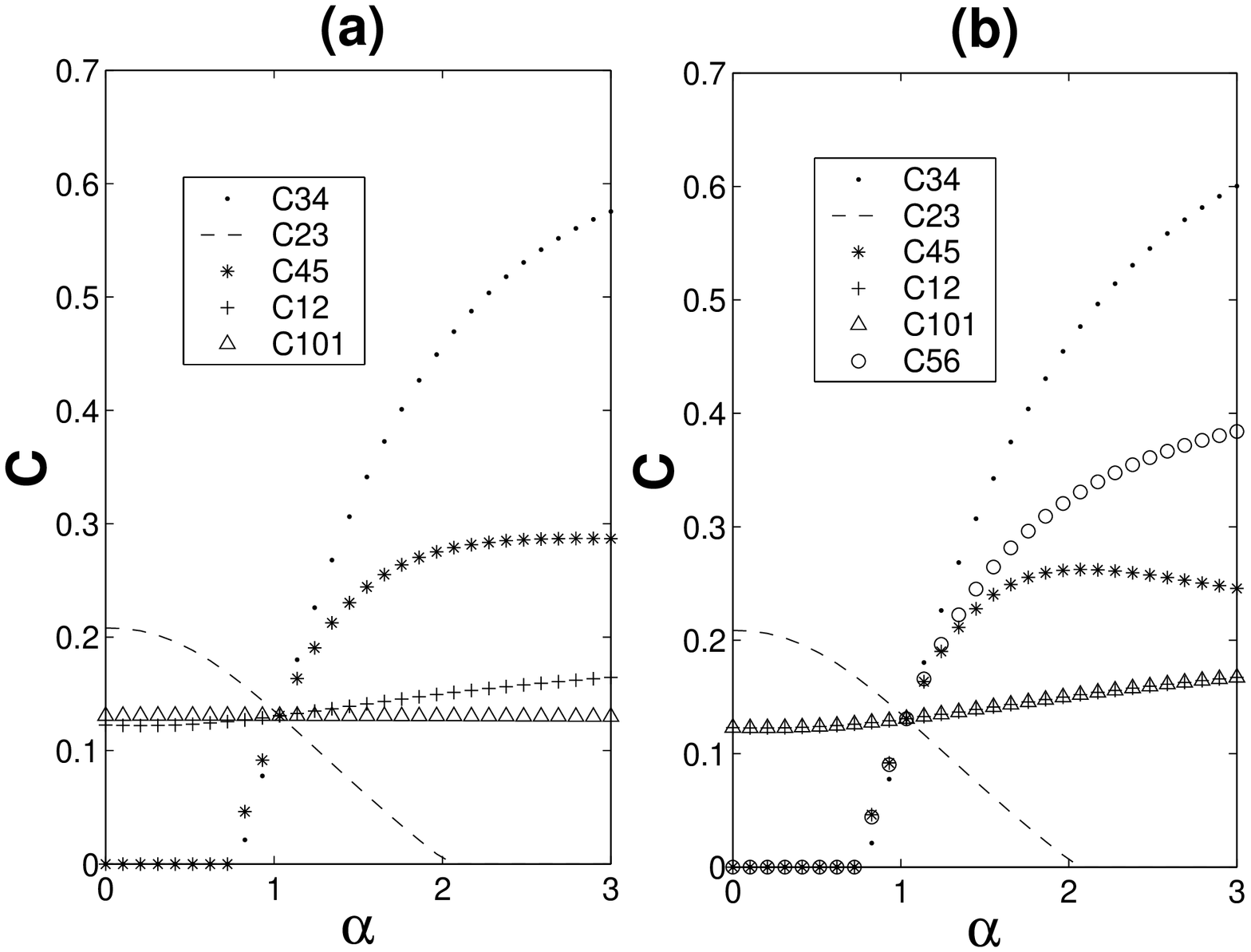}\newline
}
\caption{Nearest neighboring concurrences as a function of $\protect\alpha $
for the two-impurity model (4th and 6th are two identical impurities) (a)
and the three-impurity model (4th, 6th and 8th are three identical
impurities) (b). T=1, B=0.4, $J$=1, $J_{z}$=0.65.}
\end{figure}

Now, we review the difference between an ideal ring chain and an open chain.
For an ideal ring chain, every qubit is of the same position with the others
so that any pairwise qubits are of the same amount of entanglement. But for
an ideal open chain, pairwise entanglement is related to the position of the
qubits and exhibit oscillations due to the breaking of the symmetries. $%
^{[24]}$ 
\begin{figure}[tbph]
{\normalsize \includegraphics[width=3.5in,height=2.5in]{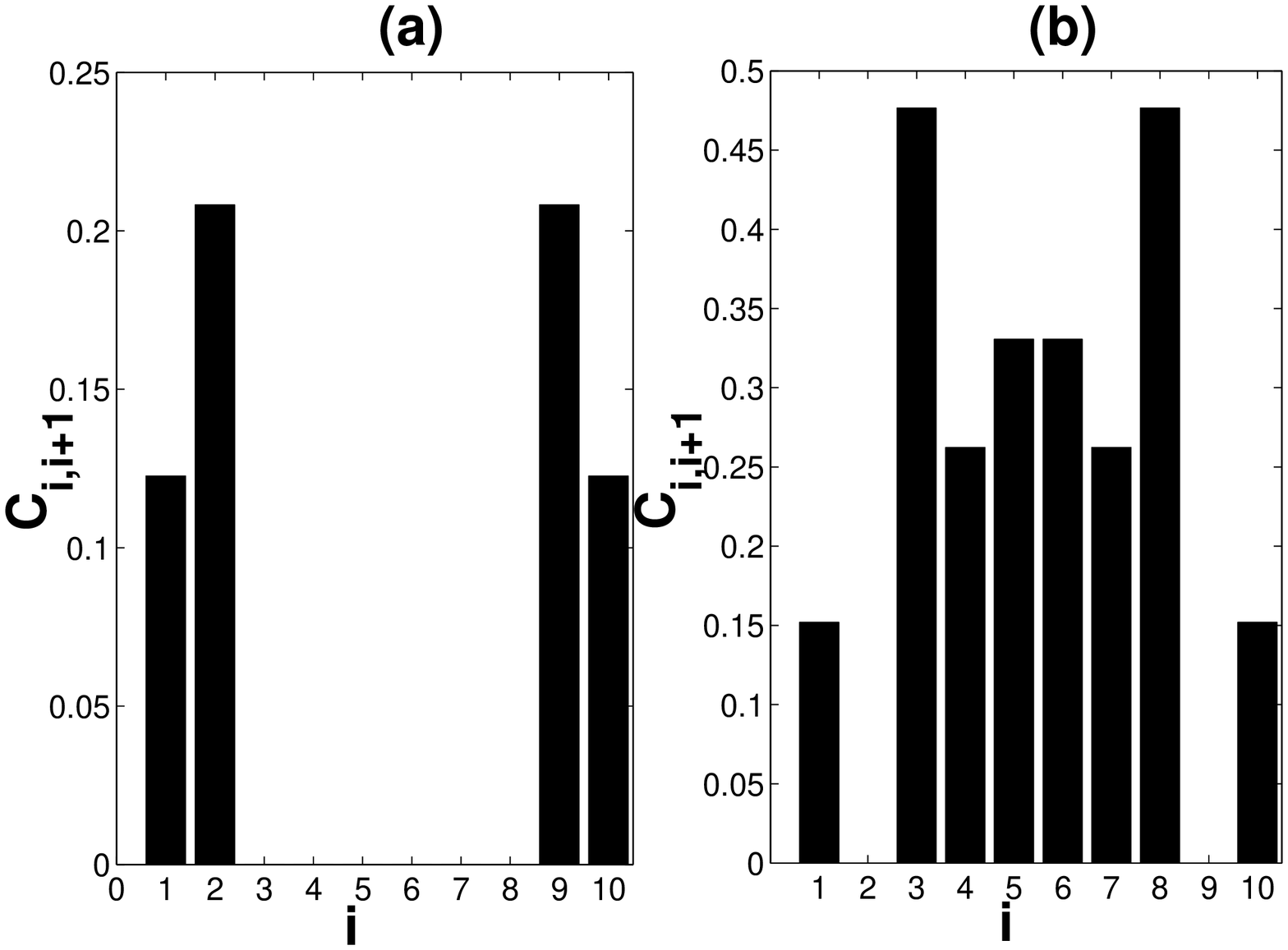}\newline
}
\caption{Nearest neighboring concurrences as a function of i for the
three-impurity model (4th, 6th and 8th are three identical impurities) $%
\protect\alpha $=0.1 (a), $\protect\alpha $=2 (b). T=1, B=0.4, $J$=1, $J_{z}$%
=0.65.}
\end{figure}

Because here we study multi-impurity, we can not obtain analytic expression
of the system. We will directly numerical calculate and plot entanglement.
In Fig. 2, we plot the pairwise entanglement as a function of $\alpha $,
corresponding to Fig. 1a and Fig. 1b, respectively. For both of the two
cases, in the regions of far away from impurities, the entanglement, for
example $C_{12}$ and $C_{101}$, are slightly affected by the various values
of $\alpha $. Within the impurities regions, we observe that there is the
almost same threshold value of $\alpha $, after which a qubit and its
nearest impurity start entangling such as $C_{34},C_{45}$ etc.. \ In Fig. 3,
we show clearly that the pairwise entanglement versus site $i$. \ If $\alpha 
$ is small shown in Fig.3a, the case equal to cutting at 4th and 8th, thus
the chain 9-10-1-2-3 is similar to the open chain$^{[24]}$ while the part
4-5-6-7-8 chain have no entanglement because of the weak couplings. If $%
\alpha >1$ such as $\alpha =2$ shown in Fig. 3b, we still can cut the chain
into two parts because $J^{\prime }>J$. Within the pure regions,
entanglement will compete while in containing impurity part pairwise
entanglement still compete each other.

\begin{figure}[tbph]
{\normalsize \includegraphics[width=3.5in,height=2.5in]{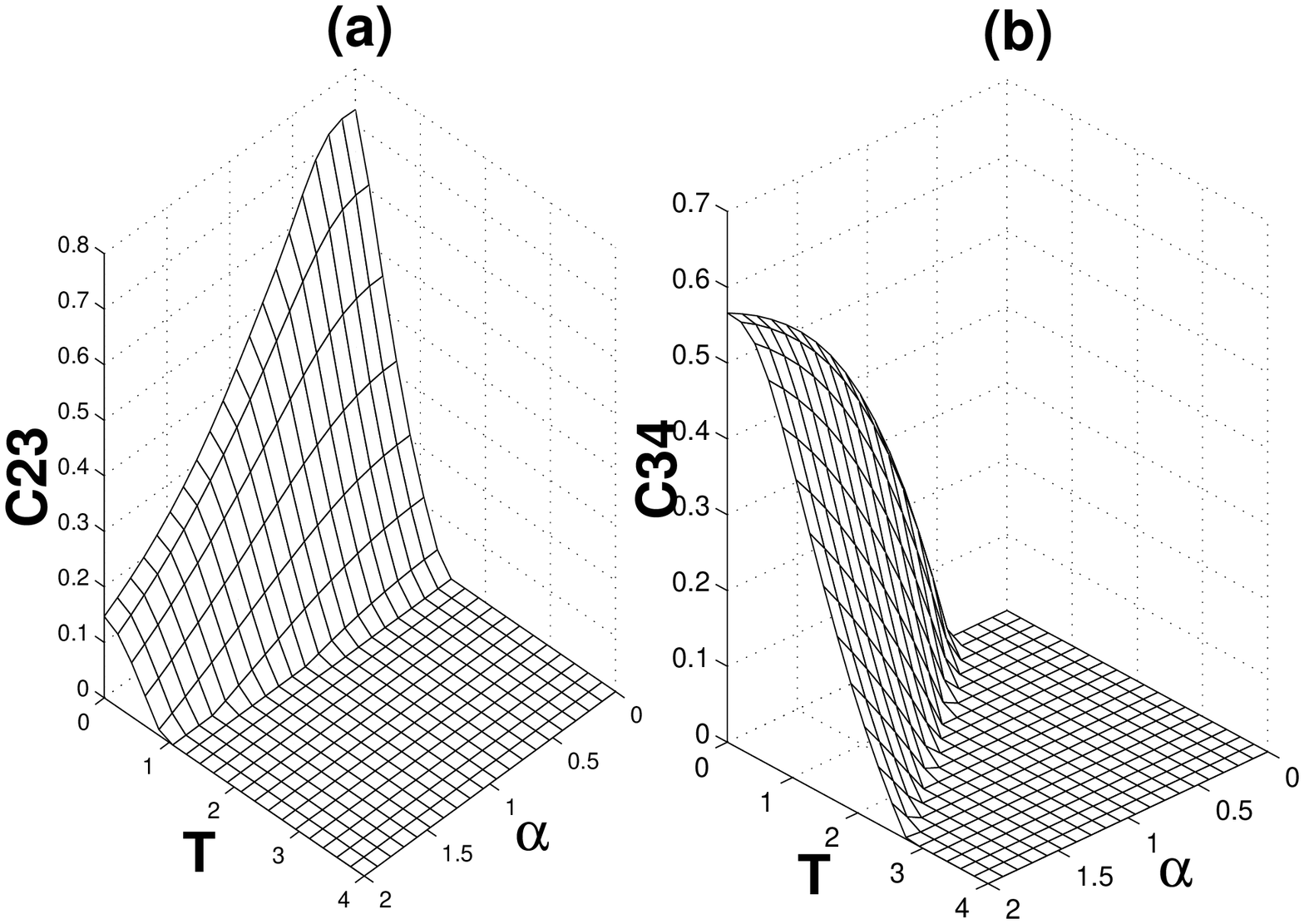}\newline
}
\caption{Nearest neighboring concurrence $C_{23}$ and $C_{34}$ versus $%
\protect\alpha $ and T for the three-impurity model (4th, 6th and 8th are
three identical impurities). $B=0.4,J=1,J_{z}=0.65.$}
\label{fig5}
\end{figure}

Fig. 4 shows the influence of temperature and the values of $\alpha $ on the
entanglement in three-impurity model. \ From this figure, we can judge again
that the second and the third qubit are pure qubits while the third and the
forth contain one impurity. Usually, it is difficult to adjust the coupling
coefficients, which means we will meet with difficult \ if we directly use
the behavior of Fig. 3 to judge which one is impurity. But we still can do
it by measuring the Concurrence changing with temperature (Refs \cite{JG,
davidavivh} proposed that Concurrence can be measured), because changing the
temperature is very easy. On the other hand, we find that $\alpha $ can
effectively enhance the Concurrence and critical temperature if $\alpha >1$
which is show in Fig. 3 and 4 clearly. By introducing impurities with large
coupling, one can also improve critical temperature and entanglement.
Therefore, our studying not only provide a stand to judge impurity but also
exhibit a way to enhance entanglement and critical temperature.

\section{Nearest Neighboring Impurity Effect}

In this section, we study the nearest neighboring impurity effect on
entanglement. \ We study the rings with a structure of Fig. 5. According to
the Fig. 5a,\FRAME{ftbpFU}{3.5466in}{2.29in}{0pt}{\Qcb{Two configurations of
spin ring with nearest-neighbor impurity. (a): qubit ring formed with 10
qubits. The 5th and 6th are two identical impurities. (b): qubit ring formed
with 10 qubits. The 4th, 7th and 8th are three identical impurities. }}{}{%
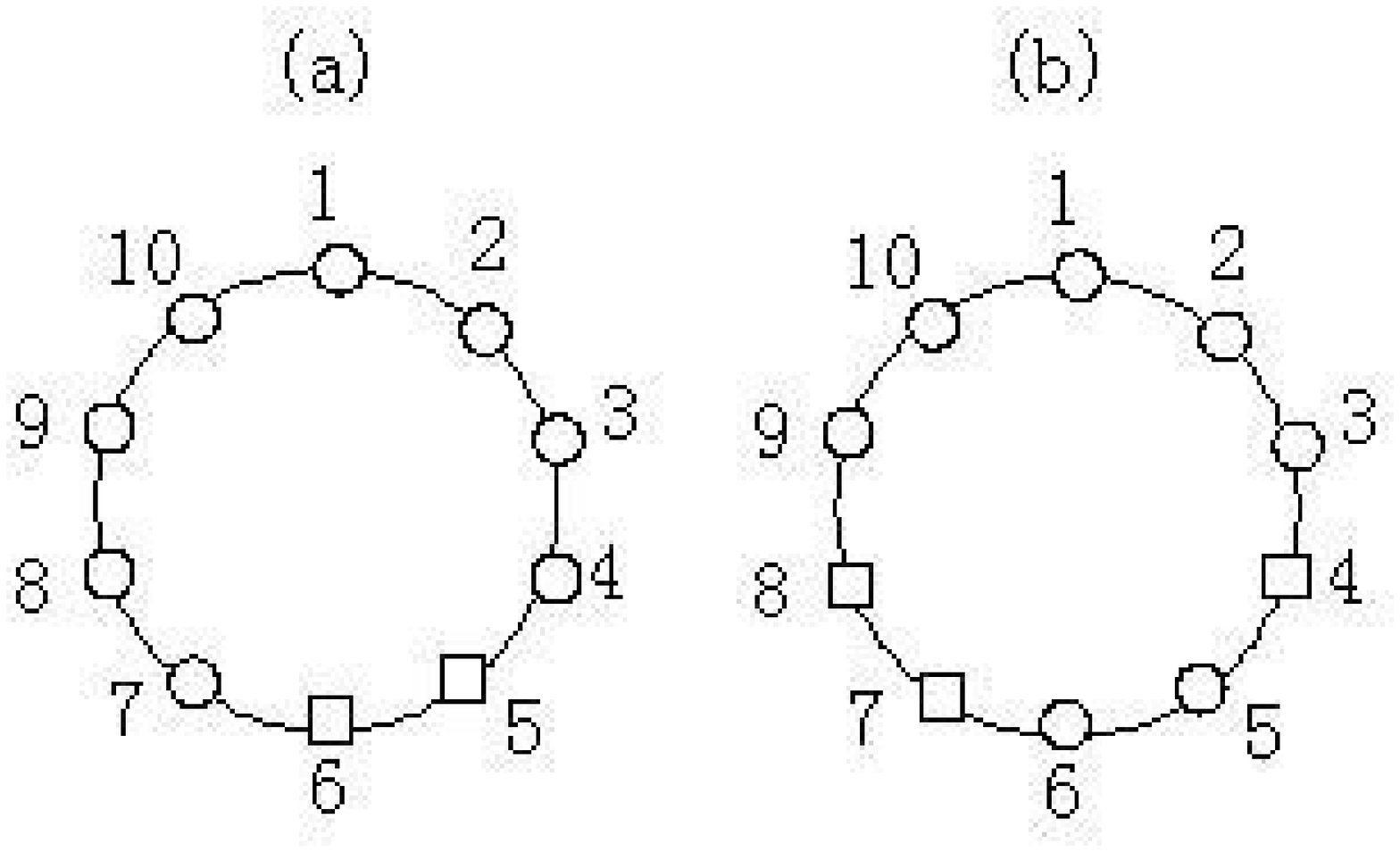}{\raisebox{-2.29in}{\includegraphics[height=2.29in]{fig5.eps}}}

\bigskip 
\begin{figure}[bp]
{\normalsize \includegraphics[width=3.5in,height=2.5in]{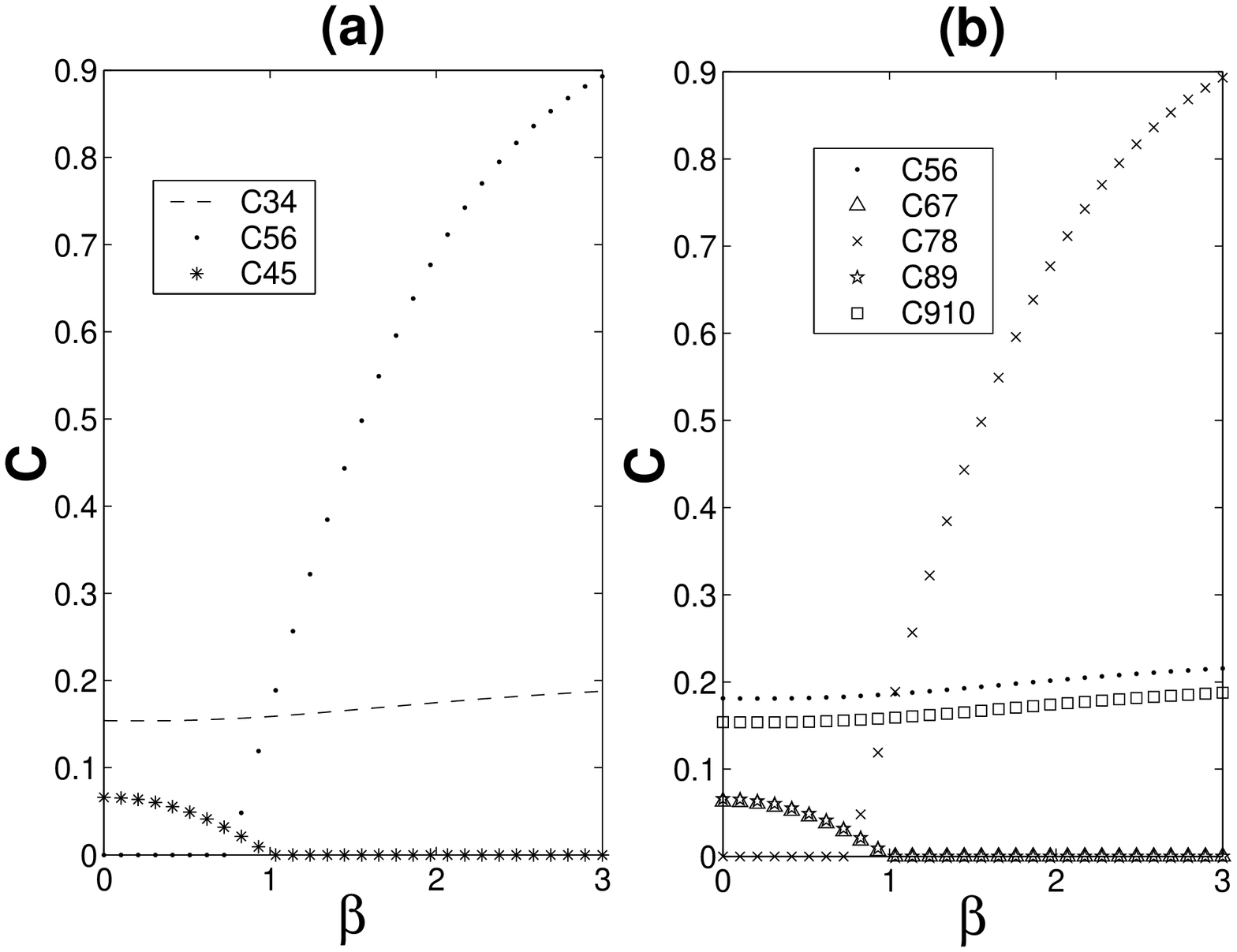} }
\caption{Nearest neighboring concurrences versus $\protect\beta $ for the
two-nearest-impurity model (5th and 6th are two identical impurities) (a)
and the three-nearest-impurity model (4th, 7th and 8th are three identical
impurities) (b). B=0.4, $J$=1, $J_{z}$=0.65, $\protect\alpha $=0.8.}
\label{fig6}
\end{figure}
the Hamiltonian is

{\normalsize 
\begin{eqnarray}
H &=&\frac{1}{2}\sum_{i=1}^{3}[J(\sigma _{i}^{x}\sigma _{i+1}^{x}+\sigma
_{i}^{y}\sigma _{i+1}^{y})+J_{z}\sigma _{i}^{z}\sigma _{i+1}^{z}]  \notag \\
&&+\frac{1}{2}\sum_{i=7}^{N}[J(\sigma _{i}^{x}\sigma _{i+1}^{x}+\sigma
_{i}^{y}\sigma _{i+1}^{y})+J_{z}\sigma _{i}^{z}\sigma _{i+1}^{z}]  \notag \\
&&+\frac{1}{2}\sum_{i=4,6}[J^{^{\prime }}(\sigma _{i}^{x}\sigma
_{i+1}^{x}+\sigma _{i}^{y}\sigma _{i+1}^{y})+J_{z}^{^{\prime }}\sigma
_{i}^{z}\sigma _{i+1}^{z}]  \notag \\
&&+\frac{1}{2}\sum_{i=5}[J^{^{\prime \prime }}(\sigma _{i}^{x}\sigma
_{i+1}^{x}+\sigma _{i}^{y}\sigma _{i+1}^{y})+J_{z}^{^{\prime \prime }}\sigma
_{i}^{z}\sigma _{i+1}^{z}]  \notag \\
&&+\frac{1}{2}\sum_{i=1}^{N}B(\sigma _{i}^{z}+\sigma _{i+1}^{z}),
\end{eqnarray}%
}

with%
\begin{equation}
J^{^{\prime \prime }}=\beta \ast J,J_{z}^{^{\prime \prime }}=\beta \ast
J_{z},
\end{equation}%
where $\beta $ characterizes the relative strength of the extra coupling
between the two nearest neighbor impurities and $J^{^{\prime }}$, $%
J_{z}^{^{\prime }}$ still has the relation of Eq.(2). Similarly, one can
write the Hamiltonian corresponding to Fig. 5b. \ We plot the pairwise
Concurrence near the two-nearest-impurity qubits area as a function of $%
\beta $ which is shown in Fig. 6. From this figure, we can see easily that
the nearest neighboring impurities coupling only affect the nearest
two-impurity and the others which couple with the impurities. For example,
in Fig.6a, the nearest neighbor impurities $C_{56}$ has a threshold value of 
$\beta $, affected by $\beta $ heavily while $C_{45}$ also will decrease as
a results of the competition between neighbor qubits. For the case of
Fig.5b, although we have more impurities, the nearest neighbor coupling only
affect entanglement of themselves $C_{78}$ and that coupling with the
impurities $C_{67}$, $C_{89}$; and all the others pairwise entanglement
almost can not be affected.

\section{Conclusion}

In conclusion, for a Heisenberg XXZ ring under a homogeneous magnetic field,
we have studied entanglement in two-impurity and three-impurity under the
two case of non-nearest-impurity and nearest-impurity. We find that the
introducing of impurities make the originally equal pairwise entanglement
compete with each other. For the weak and strong $\alpha $, we can cut the
ring chain into different open chain and then use the open chain property to
explain the competition. For the case with nearest neighbor qubits , the
change of the relative coupling $\beta $ can only affect the qubits which
couple to the impurities. If introducing impurity with large $\alpha $ and $%
\beta $, the pairwise entanglement, which couple with the impurities
directly, can be enhanced and critical temperature also will be improved.

This work was supported by Natural Science Foundation of China under Grant
No. 10575017 and Natural Science Foundation of Liaoning Province of China
under No. 20031073.

\vskip3.5mm

\section{\ References}


\begin{thebibliography}{99}
\bibitem{1} Bennett C H and DiVincenze D P Nature 2000 404 247

\bibitem{2} Nielsen M A Phys. Rev. A 2001 63 022114

\bibitem{3} Kamta G L and Starace A F Phys. Rev. Lett. 2002 88 107901

\bibitem{4} O'Connor K M and Wootters W K Phys. Rev. A 2001 63 052302

\bibitem{5} Deng L L and Man S L Chin. Phys. 2002 11 383

\bibitem{6} Yeo Y Phys. Rev. A 2002 66 062312

\bibitem{7} Sun J R, Wei Y N and Pu F K Chin. Phys. 1995 4 542

\bibitem{8} Zhang G F, Li S S and Liang J Q Opt. Commun. 2005 245 457

\bibitem{9} Zhou L, Song H S, Guo Y Q and Li C Phys. Rev. A 2003 68 024301

\bibitem{10} Wang X G Phys. Rev. A 2001 64 012313

\bibitem{11} Gunlycke D, Kendon V M and Vedral V 2001 Phys. Rev. A 64 042302

\bibitem{12} Arnesen M C, Bose S and Vedral V Phys. Rev. Lett. 2001 87 017901

\bibitem{13} Canosa N and Rossignoli R Phys. Rev. A 2004 69 052306

\bibitem{14} Khveshchenko D V Phys. Rev. B 2003 68 193307

\bibitem{15} Zhang Y M and Xu B W Chin. Phys. 1995 4 842

\bibitem{16} Zhang T, Hui X Q and Yue R H Acta Phys. Sin. 2004 53 2755

\bibitem{17} Shang Y M and Yao K L Chin. Phys. 1998 7 864

\bibitem{18} Sun Y, Chen Y G and Chen H Phys. Rev. A 2003 68 044301

\bibitem{19} Shao Y Z, Lan T and Lin G M Acta Phys. Sin. 2001 50 948

\bibitem{20} Dong Z H and Feng S P Chin. Phys. 1998 7 348

\bibitem{21} Hui X Q, Chen W X, Liu Q and Yue R H Acta Phys. Sin. 2006 55
3026

\bibitem{22} Fu H, Solomon A I and Wang X J. Phys. A 2002 35 4293

\bibitem{23} Xi X Q, Chen W X, Hao S R and Yue R H Phys. Lett. A 2002 297 291

\bibitem{24} Wang X G Phys. Rev. E 2004 69 066118

\bibitem{25} Bennett C H, DiVincenzo D P, Smolin J A and Wootters W K Phys.
Rev. A 1996 54 3824

\bibitem{26} Wootters W K Phys. Rev. Lett. 1998 80 2254

\bibitem{27} Hill S and Wootters W K Phys. Rev. Lett. 1997 78 5022

\bibitem{28} Anteneodo C and Souza A M C J. of Opt. B:Quantum Semiclassical
Opt. 2003 5 73

\bibitem{JG} Sancho J M G and Huelga S F Phys. Rev. A 2000 61 042303

\bibitem{davidavivh} Davidovich L ''Entanglement As An Observable'' 2006
reported in Texas A\&M University
\end{thebibliography}
\end{document}